\documentclass[journal,draftclsnofoot,onecolumn]{IEEEtran}

\usepackage{amssymb}

\usepackage{amsmath}

\usepackage{cite} 

\usepackage{mathrsfs}

\usepackage{graphicx}

\usepackage[square, comma, sort&compress, numbers]{natbib}

\begin{document}

\newtheorem{theorem}{Theorem}
\newtheorem{proposition}{Proposition}
\newtheorem{lemma}{Lemma}
\newtheorem{definition}{Definition}
\newtheorem{corollary}{Corollary}
\newtheorem{remark}{Remark}
\newtheorem{example}{Example}

\title{Further results on some classes of permutation polynomials over finite fields
 \thanks{ This work is supported in part by "Funding for scientific research start-up" of Nanjing Tech University.} }

\author{Xiaogang~Liu  
\thanks{
X. Liu is with
College of Computer Science and Technology,
Nanjing Tech University, 
Nanjing City,
Jiangsu Province,
PR China
211800
     e-mail:liuxg0201@163.com. }
}

\maketitle

\begin{abstract}
Let $\mathbb{F}_q$ denote the finite field with $q$ elements. The permutation behavior of several classes of infinite families of permutation polynomials over finite fields have been studied in recent years. In this paper, we continue with their studies, and get some further results about the permutation properties of the permutation polynomials. Also,  some new classes of permutation polynomials are constructed. For these, we alter the coefficients, exponents or the underlying fields, etc.

\end{abstract}


\section{Introduction}\label{secI}
 
For a  prime power $q$, let $\mathbb{F}_q$ denote the finite field of order $q$, and $\mathbb{F}_q^*$ the multiplicative  group. A polynomial $f(x)\in \mathbb{F}_q$ is called a permutation polynomial (PP)
over $\mathbb{F}_q$, if the associated polynomial mapping $f:c\rightarrow f(c)$ is a permutation of $\mathbb{F}_q$. In recent years, permutation polynomials have become an interesting area of research. They have applications in coding theroy, cryptography and combinatorial design theory \cite{FF01,V06}. Permutation polynomials with few terms attract many authors' attention for their simple algebraic structures. In particular, there are many results about permutation binomials and trinomials \cite{C1,D01,LL01,PL}.  For recent achievements on the study of permutation polynomials, the reader can consult \cite{D01,H1,LL01,L1,L2,MG1,X01}, and the references therein..

Permutation polynomials of the form $x+\mathbb{F}_q^{q^n}(x^k)$ have been studied for special $\gamma \in \mathbb{F}_{q^n}$, with even characteristic \cite{CK1,CK2}, and for the case $n=2$, i.e., trinomial permutations \cite{KZ2,LL01}. Later, Zheng, Yuan and Yu investigated permutation polynomials with the form
\[
cx-x^s+x^{qs}
\]
over $\mathbb{F}_{q^n}$, where $s$ is a positive integer, and $c\in \mathbb{F}_{q}$ \cite{ZYY}. For this, they used the well-known result (Lemma \ref{l01}) to prove three classes of permutation trinomials. For the fourth class of permutation trinomials, symbolic computation method related with resultants was used. And, they found a new relation with a class of PPs of the form
\[
(x^{q^k}-x+\delta)^s+cx
\]
over $\mathbb{F}_{q^n}$, where $\delta$ is arbitrary and $c\in \mathbb{F}_q$. This class of permutation polynomials are related to $\delta$. Based on their relationship, the aforementioned class of pemutation trinomials without restriction on $\delta$ are derived. 

    Complete permutation polynomial (CPP) is a permutation polynomial such that $f(x)+x$ is also a permutation polynomial. CPPs were introduced with the construction of Latin squares \cite{M1}. In the study of CPPs, Li etc., found that certain polynomials over $\mathbb{F}_{2^n}$ can have the same permutation properties as $ax^k+bx$ over $\mathbb{F}_{2^m}$ for positive divisor $m$ of $n$ such that $n\over m$ is odd \cite{LLLZ}. They constructed some permutation binomials, and thus more permutation polynomials of the form $a[\textup{Tr}_m^n(x)]^k+u(c+x)(\textup{Tr}_m^n(x)+x)+bx$ can be obtained, here $a,b,c,u\in \mathbb{F}_{2^m}^*$ are constants. They also studied permutation polynomials over $\mathbb{F}_{p^{2m}}$ of the form $ax^{p^m}+bx+h(x^{{p^m}} \pm x)$, and Niho-type permutation trinomials  over $\mathbb{F}_{p^{2m}}$  were constructed. 

 In this paper, we revisit the permutation polynomials proposed in \cite{ZYY,LLLZ}. Based on their results we make some modifications on the condtions, and get some further results which are generalizations. Also, we construct some new classes of  permutation polynomials in Sections \ref{Section II} and \ref{Section III}. Before going on, let us present the following lemmas, which might be useful for our study.

  \begin{lemma}\cite{ZYY}\label{l02}
Let $m,k$ be integers with $0< k < m$ and $l=\textup{gcd}(k,m)$. Let $c\in \mathbb{F}_{q^l}^*$ and $g(x)\in \mathbb{F}_{q^m}[x]$. Then $g(x^{q^k}-x+\delta)+cx$ permutes $\mathbb{F}_{q^m}$ for each $\delta\in \mathbb{F}_{q^m}$ if and only if $h(x)=g(x)^{q^k}-g(x)+cx$ permutes $\mathbb{F}_{q^m}$.

\end{lemma}

Using the same idea as in \cite[Proposition 3]{ZYY}, we can get the following lemma.
\begin{lemma}\label{l02c}
Let $m,k$ be integers with $0< k < m$ and $l=\textup{gcd}(k,m)$. Let $c\in \mathbb{F}_{q^l}^*$ and $g(x)\in \mathbb{F}_{q^m}[x]$. Then $g(x^{q^k}+x+\delta)+cx$ permutes $\mathbb{F}_{q^m}$ for each $\delta\in \mathbb{F}_{q^m}$ if and only if $h(x)=g(x)^{q^k}+g(x)+cx$ permutes $\mathbb{F}_{q^m}$.

\end{lemma}

\begin{lemma}\label{l01}\cite{Z01}
Let $d,r>0$ with $d\mid q-1$, and let $h(x)\in \mathbb{F}_q[x]$. Then
$f(x)=x^rh(x^{(q-1)/d})$ permutes $\mathbb{F}_q$ if and only if the following two conditions hold:

\begin{enumerate}
\renewcommand{\labelenumi}{$($\mbox{\roman{enumi}}$)$}
\item
$ \textup{gcd}(r,(q-1)/d)=1;$
\item
$x^rh(x)^{(q-1)/d}$ permutes $\mu_{d}, $ where $\mu_{d}$ denotes the $d$-th root of unity in $\mathbb{F}_q$.
 
\end{enumerate}

\end{lemma}

\begin{lemma}\cite{TZLH}\label{l03}
For a positive integer $m$, and $a,b\in \mathbb{F}_{2^{2m}}^*$ satisfying $\textup{Tr}_1^n({b\over {a^2}})=0$. Then the quadratic equation $x^2+ax+b=0$ has

\begin{enumerate}
\renewcommand{\labelenumi}{$($\mbox{\roman{enumi}}$)$}
\item
both solutions in the unit circle if and only if 
\begin{center}
$b={a\over {a^{2^m}}}$ and $\textup{Tr}_1^m({b\over {a^2}})=\textup{Tr}_1^m({1\over {a^{2^m+1}}})=1$;
\end{center}
\item
exactly on solution in the unit circle if and only if
\begin{center}
$b\not={a\over {a^{2^m}}}$ and $(1+b^{2^m+1})(1+a^{2^m+1}+b^{2^m+1})+a^2b^{2^m}+a^{2^{m+1}}b=0$.
 \end{center}
\end{enumerate}

\end{lemma}

\section{Permutation polynomials with two or three terms}\label{Section II}

In this section, we study the permutation behavior of four kinds of polynomials. Two kinds of them have    three terms, or  to some extent are related with a  three term polynomial, see Corollary \ref{C1} and Proposition \ref{P2}. Two kinds of them have two terms, see Proposition \ref{P03} and Proposition \ref{P04}. They come from transformations from permutation polynomials proposed in \cite{LLLZ,ZYY}. %modifications of permutation polynomials from \cite{ZYY}, others are transformations from \cite{LLLZ}.

%\subsection{PPs related with three terms}\label{SS1}

\begin{proposition}\label{P1}
For a positive integer $m$, a fixed $\delta \in \mathbb{F}_{2^{3m}}$ and $ c\in  \mathbb{F}_{2^m}^*$, the polynomial
\[
f(x)=(x^{2^m}+x+\delta)^{2^{2m}+1}+cx
\]
is a permutation of $\mathbb{F}_{q^{3m}}$.% if $s=2^{2m}+1$:

\end{proposition}

%The case  $c=1$ was presented in Proposition 4, here we consider the general  $ c\in  \mathbb{F}_{2^m}^*$.
\begin{IEEEproof}
 \cite[Proposition 4]{LWLZ}  says that
\begin{equation*}\label{e102}
g(x)=(x^{2^m}+x+\delta)^{2^{2m}+1}+ x
\end{equation*}
is a permutation of  $\mathbb{F}_{q^{3m}}$. % for every $\delta \in \mathbb{F}_{2^{3m}}$.
Set $x=c_0y, $ with $c_0\in  \mathbb{F}_{2^m}^*$. Then we have 
\begin{equation*}\label{e101}
\begin{array}{ll}
g(x)=g(c_0y)&=((c_0y)^{2^m}+ c_0y +\delta)^{2^{2m}+1}+ c_0y \\
   &={ {c_0^2}}(y^{2^m}+y+{1\over {c_0}}\delta)^{2^{2m}+1}+c_0y\\
&={ {c_0^2}}((y^{2^m}+y+{1\over {c_0}}\delta)^{2^{2m}+1}+{1\over {c_0}}y).
 \end{array}
\end{equation*}
Since $g(x)$ is a permutation for every $\delta \in \mathbb{F}_{q^{3m}}$, setting $\delta_0 ={1\over {c_0}}\delta $, 
\begin{equation*}\label{e103}
f_0(y)=(y^{2^m}+y+ \delta_0)^{2^{2m}+1}+{1\over {c_0}}y
\end{equation*}
is a permutation of $\mathbb{F}_{q^{3m}}$ for every $\delta_0 \in \mathbb{F}_{q^{3m}}$. Let $c_0={1\over c}$,  
\begin{equation*}\label{e104}
f(y)=(y^{2^m}+y+ \delta_0)^{2^{2m}+1}+cy
\end{equation*}
is a permutation of $\mathbb{F}_{q^{3m}}$ for every $\delta_0 \in \mathbb{F}_{q^{3m}}$.
\end{IEEEproof}

Note that in  Proposition \ref{P1}, $c$ is unconnected with $\delta.
$ Combing with Lemma \ref{l02},   the following result follows.
\begin{corollary}\label{C1}
For a positive integer $m$ and   $ c\in  \mathbb{F}_{2^m}^*$, the polynomial
\[
g(x)=x^{2^{m}(2^{2m}+1)}+x^{2^{2m}+1}+cx
\]
is a permutation of $\mathbb{F}_{2^{3m}}$.%  if $s=2^{2m}+1$:

\end{corollary}

\begin{example}
Set $m=3$. Using Magma, it can be verified that the following polynomial
\[
g(x)=x^{520}+x^{65}+cx
\]
is a permutation over  $\mathbb{F}_{512}$ for $c\in \mathbb{F}_8^*$.
\end{example}

In Theorem 10 \cite{ZYY}, permutation polynomials of the kind $x+x^s+x^{qs}$ are proposed, but with all coefficients $1$, here we   add a variable to the coefficient to investigate more general situations. Denote $\mu_{q+1}=\{x|x^{q+1}=1\}$. 

\begin{proposition}\label{P2}
Let $q=2^m$ be even with $q\ \equiv 1 \ (\textup{mod} \ 3)$, and $s={{q^2+q+1}\over 3}$. Then
\[
f(x)=cx+x^s+c^qx^{qs}
\]
is a permutation polynomial over $\mathbb{F}_{q^2}$ for $c\in \mathbb{F}_{q^2}^*$ satisfying $\textup{Tr}_1^m(c^{q+1})=0$.

\end{proposition}

\begin{IEEEproof}
It  can be found that $3|(q+2)$ from the assumption  $q\ \equiv 1 \ (\textup{mod} \ 3)$. Let us rewrite the polynomial $f(x)$ in the following
\begin{equation}\label{e201}
f(x)=cx+x^{{{q+2}\over 3}(q-1)+1}+c^qx^{({{q(q+2)}\over 3}+1)(q-1)+1}.
\end{equation}
Set $u={{q+2}\over 3}$, equation (\ref{e201}) becomes
\begin{equation*}\label{e202}
f(x)=x(c+x^{(q-1)u}+c^qx^{(qu+1)(q-1)}).
\end{equation*}
By Lemma \ref{l01}, $f(x)$ is a permutation polynomial if and only if
\begin{equation*}\label{e203}
g(x)=x(c+x^{u}+c^qx^{qu+1})^{q-1}
\end{equation*}
permutes $\mu_{q+1}$. %, and $\textup{gcd}()$
On the set $\mu_{q+1}$, $x^{3u}=x^{q+2}=x$ and $g(x)$ becomes
\begin{equation*}\label{e204}
g(x)=x(c+x^{u}+c^qx^{-u+1})^{q-1}.
\end{equation*}

It is necessary to show that 
\[
c+x^{u}+c^qx^{-u+1}=0
\]
has no zeros on $\mu_{q+1}$. The above equation can be written as 
\begin{equation}\label{e205}
c+x^{u}+c^qx^{2u}=0.
\end{equation}
We can find that $\textup{gcd}(u,q+1)=\textup{gcd}({{q+2}\over 3},q+1)=1$.  Set $z=x^u$, then equation (\ref{e205}) becomes
\begin{equation}\label{e206}
z^2+{1\over {c^q}}z+{c\over {c^q}}=0.
\end{equation}
For Lemma \ref{l03}, we have ${b\over {a^2}}=c^{q+1}$ which lies in $\mathbb{F}_q$. And equation (\ref{e206}) has no solutions in the unit circle by our assumption.

Let us make the following transformations for $g(x)$ on the unit circle $\mu_{q+1}$,
\begin{equation*}\label{e204}
\begin{array}{lll}
g(x)&=&x{{c^q+x^{-u}+cx^{-1+u}}\over {c+x^{u}+c^qx^{-u+1}}}\\
   &=& {{c^qx+x^{1-u}+cx^{u}}\over {c+x^{u}+c^qx^{-u+1}}}\\
 &=& {{c^qx^{3u}+x^{2u}+cx^{u}}\over {c+x^{u}+c^qx^{2u}}}\\
 &=& x^u{{c^qx^{2u}+x^{u}+c}\over {c+x^{u}+c^qx^{2u}}}\\
&=& x^u.
 \end{array}
\end{equation*}
Thus, $g(x)$ is a permutation of $\mu_{q+1}$.
\end{IEEEproof}

\begin{example}
Set $m=4$. Using Magma, it can be verified that
\[
f(x)=cx+x^{91}+c^{16}x^{1456}
\]
is a permutation polynomial over $\mathbb{F}_{256}$ for $\text{Tr}_1^4(c^{17})=0$.
\end{example}

%\subsection{PPS related with two terms}

%\backslash\{0,1\}

For the following proposition, Proposition 1 \cite{LLLZ} considered the case of odd value $m$. Here, we study the same kind of PP but with all positive values of $m$, and using totally different method. 
\begin{proposition}\label{P03}
Let  $m$ be a positive integer, $b\in \mathbb{F}_{2^{2m}}$ and $\gamma$ is  a primitive element of $\mathbb{F}_{2^{2m}}$. Then $g(x)=x^{{{2^{2m}-1}\over 3}+1}+bx$ is a permutation polynomial of $\mathbb{F}_{2^{2m}}$ if and only if 
\[
b\not= {{\gamma^{{{2^{2m}-1}\over 3}i^{\prime}}} \over  {\gamma^{3(k-k^{\prime})+i-i^{\prime}}}+1}
({\gamma^{3(k-k^{\prime})+{{2^{2m}+2}\over 3}(i-i^{\prime})}}+1)  \ \ \text{and} \ b\not= \gamma^{{{2^{2m}-1}\over 3}s },
\]
where  $0\leq i\not= i^{\prime}\leq 2, 0\leq s\leq 2$ and $1\leq k,k^{\prime}\leq {{{2^{2m}-1}\over 3}}$.
\end{proposition}

\begin{IEEEproof}
We can find that
\begin{equation*}\label{e501}
g(x)=x(x^{{{2^{2m}-1}\over 3}}+b),
\end{equation*}
and $g(0)=0$. Let us consider the nonzero elements in $\mathbb{F}_{2^{2m}}^*$.

It is not difficult to check that
\begin{equation*}\label{e502}
\mathbb{F}_{2^{2m}}^*=U_1\cup U_2 \cup U_3,
\end{equation*}
with
\begin{equation*}\label{e503}
U_i=\{\gamma^{3k+i}|1\leq k\leq {{{2^{2m}-1}\over 3}} \}
\end{equation*}
 for $i=0,1,2$. And we have
\[
U_i\cap U_j=\emptyset
\]
for $i\not=j$. Then $g(x)$ maps $U_i$ to the set
\begin{equation*}\label{e504}
V_i=\{\gamma^{3k+i}(\gamma^{{{2^{2m}-1}\over 3}i}+b)|1\leq k\leq {{{2^{2m}-1}\over 3}} \}  
\end{equation*}
for $i=0,1,2$. For fixed $i$, the elements in $V_i$ are different. Then $g(x)$ is a permutation polynomial if and only if
\begin{equation*}\label{e504}
V_i\cap V_j=\emptyset  \ \ \text{and } \ 0\notin V_s
\end{equation*}
for $i\not=j$. Which is equivalent to that
\begin{equation*}\label{505}
 \gamma^{3k+i}(\gamma^{{{2^{2m}-1}\over 3}i}+b)\not= \gamma^{3k^{\prime}+i^{\prime}}(\gamma^{{{2^{2m}-1}\over 3}i^{\prime}}+b) \ \ \text{and} \ b\not= \gamma^{{{2^{2m}-1}\over 3}s }
\end{equation*}
for $i\not= i^{\prime}$ and $1\leq k,k^{\prime}\leq {{{2^{2m}-1}\over 3}} $. After simplification, we get the result.
\end{IEEEproof}

\begin{example}
Let $m=4,$ and $\gamma$ be a primitive element of $\mathbb{F}_{256}$. For $i=2,i^{\prime}=1$, the polynomial 
\[
g(x)=x^{86}+bx
\] 
is not a permutation polynomial when $b={{\gamma^{85}}\over {\gamma^{3(k-k^{\prime})+1}+1}}( {\gamma^{3(k-k^{\prime})+86}+1})$, here $1\leq k,k^{\prime}\leq 85$.
\end{example}

\begin{proposition}\label{P04}
Let $r,i,m$ be positive integers with $\textup{gcd}(r-i,2^m+1)=1$ and $\textup{gcd}(r,i(2^m-1))=1$, and $b\in \mathbb{F}_{2^{2m}}^*$ satisfying $b^{{2^{2m}-1}\over {\textup{gcd}(i(2^m-1),2^{2m}-1)}}\not=1, b^{2^m+1}=1$. Then the polynomial 
\[
f(x)=x^{i(2^m-1)+r}+bx^r
\]
 is a permutation polynomial over $\mathbb{F}_{2^{2m}}^*$.
\end{proposition}

\begin{IEEEproof}
Let $g(x)=x^{r}(x+b)^{i(2^m-1)}$ and $S=\{x^{i(2^m-1)}| x\in \mathbb{F}_{2^{2m}}^*\}$.

\setlength{\unitlength}{0.1in}
\begin{picture}(10,10)(-15,2.5 )
\put(11,2){\vector(1,0){7}}
\put(10,10){\vector(0,-1){7}}
\put(19,10){\vector(0,-1){7}}
\put(11,10.5){\vector(1,0){7}}
\put(18.4,1.4){\makebox(2,1)[l]{S}}
\put(9.4,10.5){\makebox(2,1)[c]{$\mathbb{F}_{2^{2m}}^*$}}
\put(18.4,10.5){\makebox(2,1)[l]{$\mathbb{F}_{2^{2m}}^*$}}
\put(9.4,1.4){\makebox(2,1)[l]{S}}

\put(14.4,10.6){\makebox(2,1)[l]{f}}
\put(11.5,5.5){\makebox(2,1)[c]{$x^{i(2^m-1)}$}}
\put(20.5,5.5){\makebox(2,1)[c]{$x^{i(2^m-1)}$}}
\put(14.4,1.9){\makebox(2,1)[l]{g}}
\end{picture}

\noindent It can be verified that the above diagram is commutative. By assumption, we have $x+b\not=0$ for $x\in S$. So, $g(x)$ map $S$ to $S$. We need to check that $g(x)$ is bijective on $S$. For $x\in S$, it is not difficult to show that
\[
x^{2^m+1}=1,
\]
 which means that $x^{2^m}=x^{-1}$. Thus 
\begin{equation*}\label{e601}
g(x)=x^r{{(x^{2^m}+b^{2^m})^i}\over {(x+b)^i}}=x^r{{(x^{-1}+b^{-1})^i}\over {(x+b)^i}}={{x^{r-i}} \over {b^i}}
\end{equation*}
permutes $S$ since  $\textup{gcd}(r-i,2^m+1)=1$.

 For $x_1,x_2 \in \mathbb{F}_{2^{2m}}^*$ satisfying $x_1^{i(2^m-1)}=x_2^{i(2^m-1)}$, the following holds 
\[
({x_1\over x_2})^{i(2^m-1)}=1.
\]
 If 
\begin{equation*}\label{e701}
f(x_1)=f(x_2),
\end{equation*}
then $x_1^r=x_2^r$. Since  $\textup{gcd}(r,i(2^m-1)=1$, we have $x_1=x_2$. That is $f(x)$ is injective on the set $\{x| x^{i(2^m-1)}=z\}$ for $z\in S$. 
Using the AGW criterion \cite{AGW}, $f(x)$ is a permutation over $\mathbb{F}_{2^{2m}}^*$. 
\end{IEEEproof}

\begin{example}
Set $m=3,r=4,i=3$. Using Magma, it can be verified that the following polynomial
\[
f(x)=x^{25}+bx^4
\]
is a permutation polynomial over $\mathbb{F}_{64}$ for $b^3\not=1,b^9=1$.
\end{example}

\section{permutation polynomials  with more than three terms}\label{Section III}
In this section, we study the permutation behavior of six kinds of polynomials. They  have   more than
three terms. Proposition \ref{P5} and Proposition \ref{P6} have the property that, after some operations the complicated terms can be canceled. Proposition \ref{P7}, Proposition \ref{P8}, Proposition \ref{P9} and Proposition \ref{P10} have the property that, after certain operations the complicated exponents can be canceled. 
  They come from transformations of the permutation polynomials proposed in \cite{LLLZ}.

The following two propositions come  from the modifications of Proposition 5 and Proposition 11 \cite{LLLZ}, where they are concerned with degree $2$ extension field of $\mathbb{F}_q$. Here we are concerned with extensions of degree $3$ and $4$.

\begin{proposition}\label{P5}
Let $c\in \mathbb{F}_{q}^*$, $\delta\in \mathbb{F}_{q^3}$. Then $f(x)=g(x^{q}-x+\delta)+cx$ is a permutation polynomial if one of the following conditions holds:
\begin{enumerate}
\renewcommand{\labelenumi}{$($\mbox{\roman{enumi}}$)$}
\item
$g(x)=u(x)^{q^2}+u(x)^q+u(x)$, and $u(x)$ is a polynomial over $\mathbb{F}_{q^3}$;
 \item
 $g(x)=x^{i(q^2+q+1)}$, and $i$ is a positive integer.
\end{enumerate}
\end{proposition}

\begin{IEEEproof}
We only prove case (ii), and case (i) can be proved similarly.

Due to Lemma \ref{l02}, f(x) is a permutation polynomial if and only if
\[
h(x)=g(x)^q-g(x)+cx
\]
is a permutation of $\mathbb{F}_{q^3}$. But
\[
g(x)^q-g(x)=(x^{i(q^2+q+1)})^q-x^{i(q^2+q+1)}=0.
\]
That is $h(x)=cx$, which is a permutation of $\mathbb{F}_{q^3}$ for $c\in \mathbb{F}_q^*$.
\end{IEEEproof}

\begin{proposition}\label{P6}
Let $c\in \mathbb{F}_{q}^*$, $\delta\in \mathbb{F}_{q^4}$. Then $f(x)=g(x^{q}+x+\delta)+cx$ is a permutation polynomial if one of the following conditions holds:
\begin{enumerate}
\renewcommand{\labelenumi}{$($\mbox{\roman{enumi}}$)$}
\item
$g(x)=c_0(u(x)^{q^3}+u(x)^{q^2}+u(x)^q+u(x))$, where $u(x)$ is a polynomial over $\mathbb{F}_{q^4}$ and $c_0\in \mathbb{F}_{q^4}^*$ satisfying $c_0^{q}+c_0=0$;
 \item
 $g(x)=c_0x^{i(q^3+q^2+q+1)}$, where $i$ is a positive integer and $c_0\in \mathbb{F}_{q^4}^*$ satisfying $c_0^{q}+c_0=0$.
\end{enumerate}
\end{proposition}

\begin{IEEEproof}
We only consider case (i), and case (ii) can be proved in a similar way.

Due to Lemma \ref{l02c}, f(x) is a permutation polynomial if and only if
\[
h(x)=g(x)^q+g(x)+cx
\]
is a permutation of $\mathbb{F}_{q^4}$. But
\[
g(x)^q+g(x)=(c_0x^{i(q^3+q^2+q+1)})^q+c_0x^{i(q^3+q^2+q+1)}=(c_0+c_0^q)x^{i(q^3+q^2+q+1)}=0.
\]
That is $h(x)=cx$, which is a permutation of $\mathbb{F}_{q^4}$ for $c\in \mathbb{F}_q^*$.
\end{IEEEproof}

\begin{remark}
Note that in the above proposition $c_0\notin \mathbb{F}_q^*$ for odd characteristic.
\end{remark}

The following four propositions are from transformations of Theorem 4 \cite{LLLZ}. As in \cite{LLLZ}, Proposition \ref{P7} and Proposition \ref{P8} are concerned with degree two extension, but with different exponents and coefficients. But Proposition \ref{P9} and Proposition \ref{P10} are concerned with degree three extension, and with different exponents and coefficients. 

\begin{proposition}\label{P7}
Le $r,s,m $ be positive integers with  $s>1, \textup{gcd}(r,2^{2m}-1)=1$, and $a,\delta \in \mathbb{F}_{2^{2m}}$ satisfying  $\textup{Tr}_1^m({{a^{2^m+1}\delta^{2^m+1}}\over ({a^{2^m+1}+\delta^{2^m+1}+1})^2}) =0$. Then the polynomial
\[
f(x)=x^{r}{(x^{s(2^m-1)}+ax^{2^m-1}+\delta)}^{2^m+1}
\]
is a permutation over $\mathbb{F}_{2^{2m}}$. 
\end{proposition}

\begin{IEEEproof}
For Lemma \ref{l01}, we have $d=2^m+1$. Since $\textup{gcd}(r,2^{2m}-1)=1$, $f(x)$ is a permutation polynomial if and only if
\begin{equation}\label{e801}
g(x)=x^{r}{(x^s+ax+\delta)}^{2^{2m}-1}
\end{equation}
permutes the unit circle $\mu_{2^m+1}$. We claim that 
\[
{x^s+ax+\delta}=0
\]
has no zeros in the unit circle. Otherwise, take the $2^m$ power on both sides of the above equation 
\begin{equation*}\label{e802}
{x^{-s}+a^{2^m}x^{-1}+\delta^{2^m}}=0.
\end{equation*}
From the above two equations, we have 
\begin{equation*}\label{e803}
x^sx^{-s} =(a^{2^m}x^{-1}+\delta^{2^m})(ax+\delta)=1,
\end{equation*}
which is equivalent to
\begin{equation*}\label{e804}
(a^{2^m} +\delta^{2^m}x)(ax+\delta)=x.
\end{equation*}
After simplification, the above equation can be transformed into
\begin{equation*}\label{e805}
a\delta^{2^m}x^2+(a^{2^m+1}+\delta^{2^m+1}+1)x+a^{2^m}\delta=0.
\end{equation*}
Thus
\begin{equation*}\label{e806}
x^2+{{(a^{2^m+1}+\delta^{2^m+1}+1)}\over {a\delta^{2^m}}}x+{{a^{2^m}\delta}\over {a\delta^{2^m}}}=0.
\end{equation*}
By Lemma \ref{l03}, the above equation has no zeros in the unit circle, since $\textup{Tr}_1^m ({{a^{2^m+1}\delta^{2^m+1}}\over ({a^{2^m+1}+\delta^{2^m+1}+1})^2}) =0$.

Now, equation (\ref{e801}) becomes
\[
g(x)=x^r,
\]
which permutes $\mu_{2^m+1}$ since $\textup{gcd}(r,2^{2m}-1)=1$.
\end{IEEEproof}

\begin{example}\
Set $r=4,s=3,m=4$. Let $\omega$ be a primitive element of $\mathbb{F}_{256}$ and $\delta=\omega$. Then the polynomial
\[
f(x)=x^4(x^{45}+ax^{15}+\delta)^{17}
\]
is a permutation polynomial over  $\mathbb{F}_{256}$ when $\textup{Tr}_1^4({{a^{17}\delta^{17}}\over ({a^{17}+\delta^{17}+1})^2}) =0$.
\end{example}

Let us present the following result first before going on.

\begin{lemma} \cite{LN1} \label{ll01}
Let $q=2^k$, where $k$ is a positive integer. The quadratic equation $x^2+ux+v=0$, where $u,v \in \mathbb{F}_q$ and $u\not=0$, has roots in $\mathbb{F}_q$ if and only if $\textup{Tr}_q(v/{u^2})=0$.
\end{lemma}
 
\begin{proposition}\label{P8}
Let $r,s,m$ be positive integers with $\textup{gcd}(r,2^{2m}-1)=1$, and $a,\delta \in \mathbb{F}_{2^{m}}^*$ satisfying   $\textup{Tr}_1^m({{a^3}/ {\delta}})=1$. Then the polynomial 
\[
f(x)=x^{r }(x^{2^{m-1}(2^m+1)}+ax^{2^m+1} +\delta)^{s(2^m-1)}
\]
is a permutation polynomial over $\mathbb{F}_{2^{2m}}$.
\end{proposition}

\begin{IEEEproof}
For Lemma \ref{l01}, we have $d=2^m-1$. Since $\textup{gcd}(r,2^{2m}-1)=1$, $f(x)$ is a permutation polynomial if and only if
\begin{equation*}\label{e901}
g(x)=x^{r }(x^{2^{m-1} }+ax +\delta )^{s(2^{2m}-1) }
\end{equation*}
permutes $\mathbb{F}_{2^m}^*$. Which means that 
\begin{equation*}\label{e902}
 x^{2^{m-1} }+ax +\delta =0
\end{equation*}
has no zeros in $\mathbb{F}_{2^m}^*$. Squaring both sides of the above equation 
\[
x+a^2x^2+\delta^2=0.
\]
Which is equivalent to
\[
x^2+{1\over {a^2}}x+{{a^2}\over {\delta^2}}=0.
\]
By Lemma \ref{ll01},  the above equation has no solutions in $\mathbb{F}_{2^m}$.
\end{IEEEproof}

\begin{example}
Set $m=4,r=4,s=3$. Let $\omega$ be a primitive element of $\mathbb{F}_{256}$, and $\delta =w^{85}$. Using Magma, it can be verified that
\[
f(x)=x^{4}(x^{136}+ax^{17}+\delta)^{45}
\]
is a permutation polynomial when $\textup{Tr}_1^m({{a^3}/ {\delta}})=1$. 
\end{example}

The following lemma might be useful not only to our next proposition. 
\begin{lemma}\label{ll1}
Let $m$ be a positive integer, and $a,b\in \mathbb{F}_{2^{3m}}^*$. Then the equation
\[
x^{2^m}+ax+b=0
\]
\begin{enumerate}
\renewcommand{\labelenumi}{$($\mbox{\roman{enumi}}$)$}
\item
has at most one possible solution ${{a^{2^{2m}}b^{2^m}+a^{2^{2m}+2^m}b+b^{2^{2m}}}\over {a^{2^{2m}+2^m+1}+1}}$, when $a^{2^{2m}+2^m+1}+1\not=0$;
 \item
 has $2^m$ solutions, when $a^{2^{2m}+2^m+1}+1=0$ and ${a^{2^{2m}}b^{2^m}+a^{2^{2m}+2^m}b+b^{2^{2m}}}=0$; 
\item
has no solutions, when $a^{2^{2m}+2^m+1}+1=0$ and ${a^{2^{2m}}b^{2^m}+a^{2^{2m}+2^m}b+b^{2^{2m}}}\not=0$.
\end{enumerate}

\end{lemma}

\begin{IEEEproof}
Let $x$ be a solution, then
\begin{equation}\label{ee1}
x^{2^m}=ax+b.
\end{equation}
Taking the $2^m$ power on both sides of the above equation,
\begin{equation*}\label{ee2}
x^{2^{2m}}=a^{2^m}x^{2^m}+b^{2^m}.
\end{equation*}
Substituting equation (\ref{ee1}) into the above equation, 
\begin{equation*}\label{ee2}
x^{2^{2m}}=a^{2^m}(ax+b)+b^{2^m}.
\end{equation*}
Taking the $2^m$ power of the above equation,  
\begin{equation*}\label{ee3}
x^{2^{3m}}=a^{2^{2m}}(a^{2^m}x^{2^m}+b^{2^m})+b^{2^{2m}}.
\end{equation*}
Substituting equation (\ref{ee1}) into the above equation,  
\begin{equation*}\label{ee4}
x^{2^{3m}}=a^{2^{2m}}(a^{2^m}(ax+b)+b^{2^m})+b^{2^{2m}}.
\end{equation*}
Noting that $x^{2^{3m}}=x$, the above equation can be transformed into
\begin{equation}\label{ee5}
(a^{2^{2m}+2^m+1}+1)x={a^{2^{2m}}b^{2^m}+a^{2^{2m}+2^m}b+b^{2^{2m}}}.
\end{equation}

For case (ii), we need to show that there are $2^m$ solutions. Taking the $2^m$ power of the right side of equation (\ref{ee5}),  
\begin{equation*}\label{ee6}
ab^{2^{2m}}+a^{2^{2m}+1}b^{2^m}+b=0.
\end{equation*}
Then we can find that $x=a^{2^{2m}}b^{2^m}$ is a solution of equation (\ref{ee1}).
Let us consider the equation 
\[
c^{2^m-1}=a.
\]
Since $a^{2^{2m}+2^m+1}=1$ and $2^{3m}-1=(2^m-1)( {2^{2m}+2^m+1})$, the above equation has $2^m-1$ solutions. For every such $c$, we can find that $c+a^{2^{2m}}b^{2^m}$ is a solution of (\ref{ee1}). 
\end{IEEEproof}

\begin{proposition}\label{P9}
Let $r,m,s$ be   positive integers, and $a,b\in \mathbb{F}_{2^{3m}}^*$. Then the polynomial
\[
f(x)=x^{r}(x^{ {2^m}(2^m-1)}+ax^{2^m-1}+b)^{ s(2^{2m}+2^m+1)}
\]
\end{proposition}
is a permutation polynomial for the following two case: 
\begin{enumerate}
\renewcommand{\labelenumi}{$($\mbox{\roman{enumi}}$)$}
\item
  $a^{2^{2m}+2^m+1}+1\not=0$  and $({{a^{2^{2m}}b^{2^m}+a^{2^{2m}+2^m}b+b^{2^{2m}}}\over {a^{2^{2m}+2^m+1}+1}})^{2^{2m}+2^m+1}\not=1$; 
\item
  $a^{2^{2m}+2^m+1}+1=0$ and ${a^{2^{2m}}b^{2^m}+a^{2^{2m}+2^m}b+b^{2^{2m}}}\not=0$.
\end{enumerate}
here $\textup{gcd}(r,2^{3m}-1)=1$.

\begin{IEEEproof}
For Lemma \ref{l01}, we have $d=2^{2m}+2^m+1$. Since  $\textup{gcd}(r,2^{3m}-1))=1$, then $f(x)$ is a permutation polynomial if and only if
\begin{equation*}\label{eee1}
g(x)=x^{r}(x^{ {2^m} }+ax +b)^{ (2^m-1)(2^{2m}+2^m+1)}=x^{r}(x^{ {2^m} }+ax +b)^{ s(2^{3m}-1)}
\end{equation*}
permutes $\mu_{2^{2m}+2^m+1}$.  We only need to show that $(x^{ {2^m} }+ax +b)^{ 2^{3m}-1}$ is nonzero on the set $\mu_{2^{2m}+2^m+1}$, which can be deduced from Lemma \ref{ll1}.
\end{IEEEproof}

\begin{example}
Let $r=4,m=3,s=3,b=1$. Using Magma, it can be verified that the following polynomial
\[
f(x)=x^4(x^{56}+ax^7+1)^{219}
\]
is a permutation polynomial over $\mathbb{F}_{2^9}$ when $a^{73}\not=1$  and $({{{a^{64}} +a^{72}+1}\over {a^{73}}})^{73}\not=1$.

\end{example}

 Before going on, let us introduce some notations first. 
Let $\mathbb{F}_{q^n}$ be the finite field which is an extension of $\mathbb{F}_{q}$ of degree $n$. 
Use $N$ to denote the norm function from $\mathbb{F}_{q^n}$ to $\mathbb{F}_q$, i.e., for $x\in \mathbb{F}_{q^n}$, $N(x)=x^{1+q+q^2+\cdots+q^{n-1}}$. Let $L(x)=ax+bx^q+x^{q^2}$ be a linearized polynomial, with $a,b\in \mathbb{F}_{q^n}^*$. Define 
\begin{equation}\label{ee05}
u={{a^q}\over {b^{q+1}}}.
\end{equation}
 There is the following result about the number of solutions of $L(x)$ over $\mathbb{F}_{q^3}$.   

%[Theorem 9]

\begin{lemma}\cite{GJ01}\label{ll2}
Let   $L(x)=ax+bx^q+x^{q^2}$ with $a,b\in \mathbb{F}_{q^3}^*$. Then $L(x)$ has $1$ root in $\mathbb{F}_{q^3}$
if and only if $1+N(b)(u^{q^2}+u^q+u+1)+N(a)\not=0$.
\end{lemma}

\begin{proposition}\label{P10}
Let $r,s,m$ be positive integers satisfying $\textup{gcd}(r,2^{3m}-1)=1, \textup{gcd}(3, 2^m-1)=1$. Then the polynomial
\[
f(x)=x^r(x^{2^{2m}(2^m-1)}+bx^{2^m(2^m-1)}+ax^{2^m-1}+\delta)^{s(2^{2m}+2^m+1)}
\]
is a permutation polynomial over $\mathbb{F}_{2^{3m}}$, where $ a+b+1\not=0, {{\delta +1} \over {a+b+1}} \in \mathbb{F}_{2^{m}}^*$, $a+b+\delta+1\not=0$ and $1+N(b)(u^{q^2}+u^q+u+1)+N(a)\not=0$. Here $u$ is defined as in (\ref{ee05}).
\end{proposition}

\begin{IEEEproof}
For Lemma \ref{l01}, we have $d=2^{2m}+2^m+1$. Since $\textup{gcd}(r,2^{3m}-1)=1$, $f(x)$ is a permutation poynomial over $\mathbb{F}_{2^{3m}}$ if and only if 
\begin{equation}\label{ee01}
g(x)=x^r(x^{2^{2m} }+bx^{2^m }+ax +\delta)^{s(2^{3m}-1)}
\end{equation}
permutes $\mu_{2^{2m}+2^m+1}$. Which is equivalent to 
\begin{equation}\label{ee02}
x^{2^{2m} }+bx^{2^m }+ax +\delta=0
\end{equation}
has no solutions in $\mu_{2^{2m}+2^m+1}$.

Since $ a+b+1\not=0, {{\delta +1} \over {a+b+1}} \in \mathbb{F}_{2^{m}}^*$, the above equation has a solution $x_0={{\delta +1} \over {a+b+1}}$ in $\mathbb{F}_{2^{m}}^*$. But $\textup{gcd}(3, 2^m-1)=1$, $x_0$ lies in  $\mu_{2^{2m}+2^m+1}$ if and only $x_0=1$. Contradiction with the assumption that $a+b+\delta+1\not=0$. 

 By Lemma \ref{ll2}, equation (\ref{ee02}) is in fact a permutation polynomial over $\mathbb{F}_{2^{3m}}$. It has only one solution $x_0$, but which does not belong to the set $\mu_{2^{2m}+2^m+1}$.
\end{IEEEproof}

\begin{example}
Set $q=4, b=1, r=4,s=3,\delta=0$. Then the two values of $a$  that satisfy our condition are $\omega^{21}$ and $\omega^{42}$ where $\omega $ is a primitive element of $\mathbb{F}_{64}$. Using Magma, it can be verified that for those two values
\[
f(x)=x^6(x^{48}+x^{12}+ax)^{63} 
\]
is a permutation polynomial over $\mathbb{F}_{64}$.
\end{example}

\section{Conclusion}

In this paper, based on the results obtained recently about permutation polynomials, we get some further PPs. Also, with suitable modifications on the condititions,  some new classes of PPs are proposed. To do this, some coefficients of the original permutaion polynomials are made to be variable, or the exponents are unfixed, or the field  extensions are different from their work.

 \section*{Acknowledgment}

 The author would like to thank the anonymous referees for helpful suggestions and comments.

%........‘On the values of representation functions’.....

\end{document}